\documentclass[10pt]{article}
\usepackage{epsfig,graphicx}
\usepackage{times,ch2005}

\def\beq{\begin{equation}}
\def\eeq#1{\label{#1}\end{equation}}
\def\eeqn{\end{equation}}

\def\beqa{\begin{eqnarray}}
\def\eeqa#1{\label{#1}\end{eqnarray}}
\def\eeqan{\end{eqnarray}}

\def\Dslash{\not{\hbox{\kern-4pt $D$}}}
\def\dslash{\not{\hbox{\kern-2pt $\del$}}}

\newcommand{\tev}{\ensuremath{\mathrm{\,Te\kern -0.1em V}}\xspace}
\newcommand{\gev}{\ensuremath{\mathrm{\,Ge\kern -0.1em V}}\xspace}
\newcommand{\mev}{\ensuremath{\mathrm{\,Me\kern -0.1em V}}\xspace}
\newcommand{\kev}{\ensuremath{\mathrm{\,ke\kern -0.1em V}}\xspace}
\newcommand{\ev}{\ensuremath{\mathrm{\,e\kern -0.1em V}}\xspace}
\newcommand{\gevc}{\ensuremath{{\mathrm{\,Ge\kern -0.1em V\!/}c}}\xspace}
\newcommand{\mevc}{\ensuremath{{\mathrm{\,Me\kern -0.1em V\!/}c}}\xspace}
\newcommand{\gevcc}{\ensuremath{{\mathrm{\,Ge\kern -0.1em V\!/}c^2}}\xspace}
\newcommand{\mevcc}{\ensuremath{{\mathrm{\,Me\kern -0.1em V\!/}c^2}}\xspace}

\def\mus  {\ensuremath{\rm \,\mus}\xspace}

\def\mus        {\ensuremath{\,\mu{\rm s}}\xspace}    

\begin{document}


\Title{Performance Limits for Cherenkov Instruments}
\bigskip


%
\label{Hofmann1Start}

%
\author{ Werner Hofmann\index{Hofmann, W.} }

%
\address{Max-Planck-Institut f\"ur Kernphysik\\
Postf. 103980 \\
D 69029 Heidelberg, Germany\\
}

\makeauthor\abstracts{
The performance of Cherenkov instruments for the detection
of very high energy gamma rays is ultimately limited by the
fluctuations in the development of air showers. With particular
emphasis on the angular resolution, the ultimate performance
limits are investigated on the basis of simulations.}

\section{Introduction}

Imaging Cherenkov telescopes and in particular stereoscopic arrays of Cherenkov
telescopes such as HEGRA or H.E.S.S. have emerged as the prime instrument
for gamma-ray astronomy in the TeV energy regime. H.E.S.S., for example, 
reaches an energy threshold around 100~GeV, an angular resolution for 
individual gamma rays of $0.1^\circ$ and a sensitivity of less then 1\% of the Crab
flux for 25~h of exposure. Considering the sensitivity as a function of energy,
one finds three different regimes. At low energy, the detection threshold is basically
determined by the number of Cherenkov photons detected, i.e. by the product of
mirror area and photodetection efficiency; typically, about 50 photoelectrons are 
required to trigger a telescope and to generate a reconstructible image. At
the high energy end, instruments are limited by the number of detected gamma rays, i.e.
by the area over which showers are detected. In the intermediate energy range, 
background from hadron- and electron-induced cosmic ray showers limits the sensitivity.
For point sources, the number of background events
under a signal is proportional to the square of
the angular resolution times the rejection efficiency for cosmic rays; an improved
angular resolution is therefore a key element in increasing the sensitivity of
instruments. Angular resolution is also of significant interest to unravel
the processes in extended gamma-ray sources such as supernova remnants. Recent H.E.S.S. measurements
have for the first time resolved the structure of supernova remnants in very
high energy gamma rays \cite{nature1713,velajr}; however, Chandra and XMM X-ray images of remnants show
much finer structures than currently resolved by Cherenkov instruments
\cite{chandra1006,chandra1713,xmm1713}. The ability
to detect such structures at TeV energies would help in identifying the origin of
the gamma rays: Inverse Compton scattering of electrons should show narrow structures comparable
to the X-ray structures -- governed by the rapid cooling of the radiating electrons --
whereas hadronic interactions are expected to generate much
smoother structures \cite{snrmodel}.

Designing future Cherenkov instruments, a key question is how much improvement is 
possible concerning angular resolution and background rejection; if current
instruments are already close to
the fundamental limitations imposed by shower physics, there is no point in pushing, e.g., 
for smaller pixel size in the photon detectors. Here, we address this question on the basis
of Monte Carlo simulations, and proceed in two steps: in a first step, we consider the ultimate
performance which could be obtained if ALL Cherenkov photons were detected on the ground and
their direction and impact point measured exactly. In a second step, we study how strongly
the performance deteriorates compared to the ideal case, if measurement errors on photon direction and 
impact point are introduced (equivalent to a finite pixel size of the photon detector and a finite
size of a telescope mirror, respectively), and if only a certain fraction of all photons is detected.
Since the algorithms used to derive gamma-ray directions from the detected photons may not be
optimal, the resulting resolutions should be considered as limits; the ultimate performance
of instruments might still be better.

\section{Angular resolution}

To study the angular resolution, CORSIKA \cite{corsika} was used to generate showers and
simulate Cherenkov emission. 
The simulations assume that a fraction $\epsilon$ of the Cherenkov photons reaching the
ground are detected and their impact point and direction is measured. The shower direction 
is then determined from a likelihood fit using a Monte-Carlo determined distribution function
$\rho(r,\theta_{par},\theta_{perp})$ where $r$ is the distance to the shower axis, 
$\theta_{par}$ the inclination of the photon direction relative to the shower direction
in the plane defined by the shower axis and the photon impact point, and $\theta_{perp}$ 
the inclination in the perpendicular plane. Shower direction and impact point are varied
to (numerically) maximize the likelihood function. This fitting procedure is certainly not 
optimal since (a) it neglects the strong correlations between Cherenkov photons, (b) approximations
were used to represent the three-dimensional distribution function and (c) the same distribution
function was used for all shower energies. Also, to limit storage and CPU time requirements, only
photons within $3^\circ$ around the assumed shower axis were used; however, most of the information
seems to be contained within the innermost $1^\circ$. Other algorithms were tried to reconstruct
the shower axis, e.g. by backtracking detected photons and searching for peaks in photon density
at different heights; the results were always comparable to the likelihood fit, or worse.

\begin{figure}[htb]
\begin{center}
\includegraphics[height=7cm]{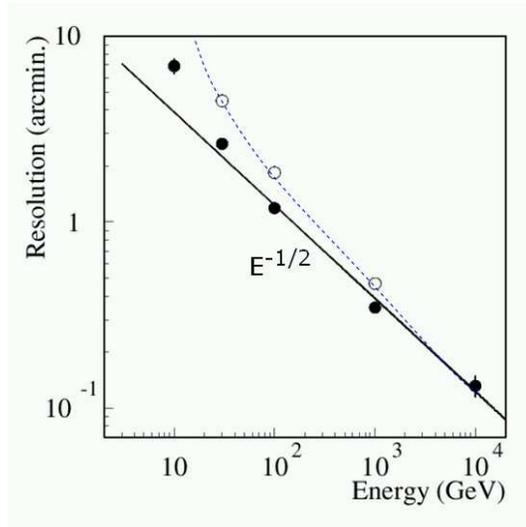}
\caption{Angular resolution obtained if all Cherenkov photons are detected, as a function of 
gamma ray energy. Closed points denote the case without geomagnetic field, or in the direction not
influenced by the field; open points show the resolution in the direction where the shower is
spread out by the field. The simulations assume near-vertical incidence at the H.E.S.S. site in
Namibia.}
\label{fig:hofmann1-fig1}
\end{center}
\end{figure}

Fig. \ref{fig:hofmann1-fig1} shows the limiting angular resolution as a function of energy. Presumably
as a result of shower fluctuations, the resolution improves like $1/\sqrt{E}$ for energies above 
30~GeV, with values around 1' at 100~GeV and 0.3' at 1~TeV. (Here and in the following, 
angular resolution is defined at the Gaussian width of the distribution of reconstructed gamma-ray
directions, projected onto one axis of the coordinate system.) The lowest point at 10~GeV deviates
from the $1/\sqrt{E}$-dependence, probably reflecting the fact that at this energy the fit frequently
did not
converge well. The full data points in Fig. \ref{fig:hofmann1-fig1}
show the resolution obtained without geomagnetic field, which
proved equivalent to the resolution obtained with geomagnetic field in the plane where shower particles are not
deflected (the simulations assume near-vertical incidence and magnetic fields for the H.E.S.S. site
in Namibia). In the direction where the field acts, resolutions at the lower energies degrade by
a factor two (at 30 GeV); at and above 1~TeV, the influence is modest. The limiting resolutions obtained
in this way are at least a factor 3 to 5 better than what is obtained with telescopes such at H.E.S.S. at 100~GeV,
and an order of magnitude better at multi-TeV energies, demonstrating that there is 
significant margin for improvement.

\begin{figure}[htb]
\begin{center}
\includegraphics[height=7cm]{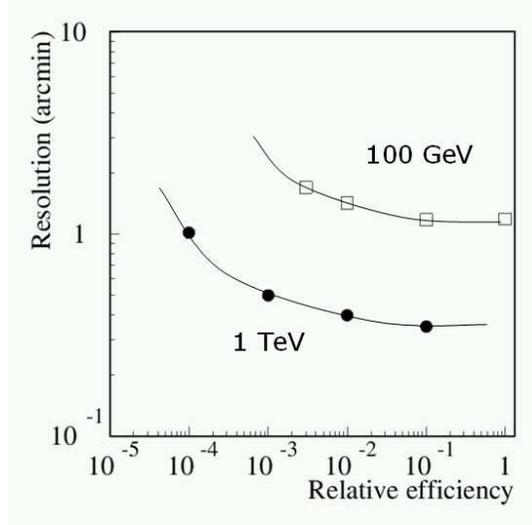}
\caption{Angular resolution at 100~GeV and at 1~TeV as a function of the detection efficiency for Cherenkov photons.
For easier comparison, efficiency is normalized to the efficiency obtained when the ground is fully covered with
current-generation Cherenkov telescopes equipped with PMT photon detectors.}
\label{fig:hofmann1-fig2}
\end{center}
\end{figure}

\begin{figure}[htb]
\begin{center}
\includegraphics[height=5.5cm]{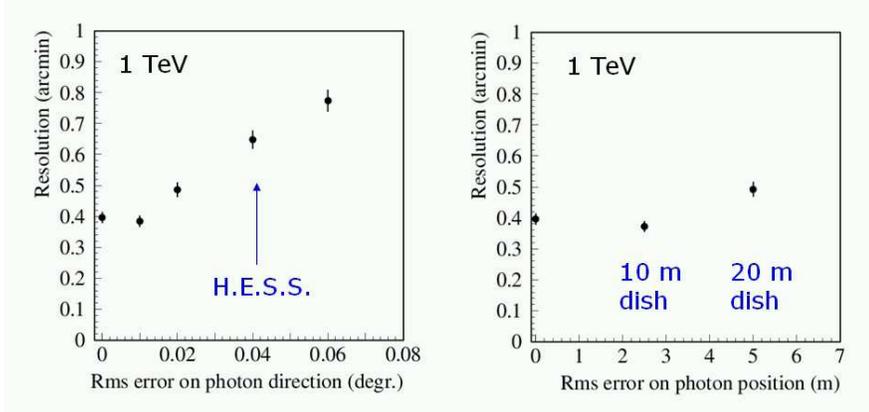}
\caption{Influence of the measurement errors on Cherenkov photon direction and 
impact point on the angular resolution. For round pixels or dish, the rms error is 1/4
of the pixel or dish size.}
\label{fig:hofmann1-fig3}
\end{center}
\end{figure}

Perfect detection of all Cherenkov photons is of course an unrealistic requirement; a system such as
H.E.S.S. with about 100~m$^2$ mirror area per telescope, 120~m telescope spacing and maybe 10\% photon detection
efficiency (in the 300 to 600 nm range) detects about $10^{-3}$ of the photons. With closer telescope spacing
(and a correspondingly larger number of telescopes) as well as improved photon detectors this number 
might be brought to $10^{-2}$. Fig. \ref{fig:hofmann1-fig2} illustrates how the angular resolution deteriorates as a 
smaller fraction of photons is detected; the assumed detection efficiency is given relative to the
detection efficiency achieved with current PMTs. For relative efficiencies about $10^{-2}$, the angular
resolution is close to the limiting resolution; for lower efficiencies, photon statistics starts to be an issue
and the resolution degrades. Fig.~\ref{fig:hofmann1-fig3} demonstrates the effect of finite pixel and mirror sizes by
introducing a Gaussian measurement error on photon direction and photon impact point (a round pixel or mirror
results in an rms measurement error of 1/4 of the pixel or mirror diameter). For the 1~TeV showers studied
here, measurement errors below $0.02^\circ$ ($0.08^\circ$ pixel size) seem to be required for best performance;
impact errors up to about 3~m (i.e., at 12~m dish) seem acceptable.

\section{Background rejection}

With a similar approach, the background rejection was studied, comparing gamma-ray induced showers with
proton-induced showers at a higher energy where they generate about the same number of 
Cherenkov photons. First results
show that at 1~TeV photon energy, a background suppression of a few $10^{-4}$ seems possible, deteriorating
to about $10^{-2}$ at 100~GeV. The dependence on detection efficiency and detector resolution remains to be
studied.

\begin{figure}[htb]
\begin{center}
\includegraphics[height=7cm]{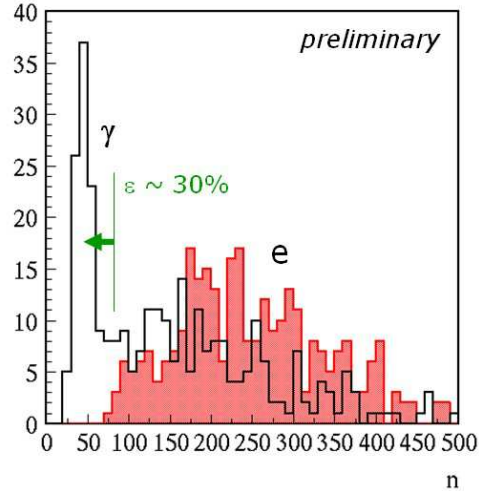}
\caption{Distribution in the number 
of photons intersecting a 10~m radius area at 17~km height above ground, for 1~TeV gamma-rays showers and
electron-induced showers, including photons from the night-sky background, for a photon detection efficiency of
$10^{-2}$.}
\label{fig:hofmann1-fig4}
\end{center}
\end{figure}

At such rejection levels, cosmic-ray electrons start dominating the background and a key question is if there
is a way to distinguish electrons and gamma rays, both generating electromagnetic showers. Basically, there
are two approaches: (a) one can used the fact that the (average) maximum of a gamma-induced shower is about one radiation
length deeper in the atmosphere than the (average) maximum of electron-induced showers, and (b) one can try to
detect the track of primary electron. 
Backtracking the Cherenkov photons, 
the height of the shower maximum can be reconstructed with a precision
of better than one radiation length; however, given the fluctuations of the photon conversion point of about
one radiation length, the difference between electrons and gamma rays will never be enough for an efficient 
separation. Reconstruction of the primary electron track was tried as follows: the photon distribution in
17~km height above ground was determined by backtracking the detected photons and all photons within 10~m
radius from the (reconstructed) shower axis were counted. Fig.~\ref{fig:hofmann1-fig4} shows the resulting distributions,
assuming a detection efficiency of $10^{-2}$ and adding a realistic amount of night-sky background noise.
The electron-induced showers are characterized by a much larger number of photons, mostly 
associated with the primary electron. If a low
gamma-ray selection efficiency of 30\% is accepted, a relatively clean gamma-ray sample can be obtained. Being able to
backtrack a photon with an error well below 10~m obviously implies very small pixels, of $0.01^\circ$ or less.

\section{The bottom line}

Financial considerations aside, shower physics seems to allow further improvement of the performance of Cherenkov
instruments in particular in the domain around a TeV and above. An ideal detector, covering a large fraction of
the ground with 10~m class Cherenkov telescopes equipped with very fine pixels could provide a gain of up to an order
of magnitude in angular resolution and in proton rejection, and non-negligible electron rejection, corresponding to
a Q-factor of about 3-5. At lower energies, shower fluctuations become more and more important and gains are
reduced to factors of a few at 100 GeV, and may be negligible at even lower energies.

In particular for the study of extended sources, there is a clear physics case for improved resolution at very
high energies, and with the advent of highly efficient semiconductor photon detectors, one may even be able to
realize such instruments at acceptable cost, using tiny pixels with a binary (photon-counting) readout. With a boost
of a factor 3 in quantum efficiency and 50~m spacing of 10-12~m-size telescopes, a sufficient number of Cherenkov
photons is detected.
An array would
still, however, need to consist of a significant number (several tens) 
of telescopes, to detect a sufficient number of gamma rays. Provided that telescope information can be
combined already at the trigger level -- which should be feasible in case of binary readout, where 
data are easily buffered - a dense array of medium-size telescopes could also be used to detect
very low-energy showers, replacing a single large dish and thereby generating an extremely versatile
instrument.

%
\label{Hofmann1End}
 
\end{document}